\documentstyle[12pt]{article}
\newcommand{\rf}[1]{(\ref{#1})}
\newcommand{\bea}{\begin{eqnarray}}
\newcommand{\eea}{\end{eqnarray}}
\newcommand{\g}{\gamma}

\renewcommand{\a}{\alpha}

\newcommand{\m}{\mu}

\newcommand{\k}{\kappa}

\newcommand{\ra}{\right\rangle}
\newcommand{\la}{\left\langle}

\newcommand{\cT}{{\cal T}}
\newcommand{\cN}{{\cal N}}

\newcommand{\cO}{{\cal O}}

\def\void{}
\def\labelmark{}

\newenvironment{formula}[1]{\def\labelname{#1}
\ifx\void\labelname\def\junk{\begin{displaymath}}
\else\def\junk{\begin{equation}\label{\labelname}}\fi\junk}%
{\ifx\void\labelname\def\junk{\end{displaymath}}
\else\def\junk{\end{equation}}\fi\junk\labelmark\def\labelname{}}

{\ifx\void\labelname\def\junk{\end{array}\end{displaymath}}
\else\def\junk{\end{array}\right.\end{equation}}
\fi\junk\labelmark\def\labelname{}\def\junk{}
}

\newcommand{\beq}{\begin{formula}}
\newcommand{\eeq}{\end{formula}}
\newcommand{\beqv}{\begin{formula}{}}

\begin{document}
\topmargin 0pt
\oddsidemargin 5mm
\headheight 0pt
\headsep 0pt
\topskip 9mm

\hfill    NBI-HE-93-2

\hfill January 1993

\begin{center}
\vspace{24pt}
{\large \bf Observing 4d  baby universes in quantum gravity}

\vspace{24pt}

{\sl J. Ambj\o rn }

\vspace{6pt}

 The Niels Bohr Institute\\
Blegdamsvej 17, DK-2100 Copenhagen \O , Denmark\\

\vspace{12pt}

{\sl S. Jain}

\vspace{6pt}

Centre for Theoretical Studies, Indian Institute of Science\\
Bangalore 560012, India\\

\vspace{12pt}

{\sl J. Jurkiewicz}{\footnote {Supported by the
KBN grant no. 2 0053 91 01}}

\vspace{6pt}

Inst. of Phys., Jagellonian University., \\
ul. Reymonta 4, PL-30 059, Krak\'{o}w~16, Poland

\vspace{12pt}

{\sl C.F. Kristjansen}

\vspace{6pt}

 The Niels Bohr Institute\\
Blegdamsvej 17, DK-2100 Copenhagen \O , Denmark\\

\end{center}

\vspace{24pt}

\addtolength{\baselineskip}{0.20\baselineskip}
\vfill

\begin{center}
{\bf Abstract}
\end{center}

\vspace{12pt}

\noindent
We measure the fractal structure of four dimensional simplicial quantum
gravity by identifying so-called baby universes.
This allows an easy  determination of the critical
exponent $\g$ connected to the entropy of four-dimensional manifolds.
%\vspace{24pt}

\vfill

\newpage

\section{Introduction}

A  simple way to characterize the fractal structure
of 2d quantum gravity  was recently introduced in ref.\ \cite{jm}.
Using very few assumptions it was possible to prove that the
functional behaviour of the number of
two-dimensional surfaces with area $N_2$ and
coupled to matter with central charge $c <1$
\beq{*1}
\cN (N_2) \sim N_{2}^{\g(c) -3}\; e^{\m_0 N_2}
\eeq
is directly related to the fractal structure of 2d quantum gravity.
In particular one can determine $\g(c)$ by simply counting the number
of so-called ``minimum bottleneck universes'', abbreviated ``minbu's'',
living on a typical surface in the ensemble of surfaces
of fixed total area. This number is given by
\beq{*2}
\cN_{N_2} (V_2) \sim N_2 V_2^{\g(c) -2},
\eeq
where $N_2$ is the total area of the surface
and $V_2 \ll N_2$ is the area of the minbu's being counted.
A "baby universe" of area $V_2$ is any simply connected
region of the surface of area $V_2$ and boundary length
$l$ such that $V_2 \gg l^2$, and a minbu is a baby universe
whose $l$ is the smallest possible length
consistent with the ultraviolet cutoff. This boundary,
along which the minbu is connected to the remainder
of the (parent) surface is called a minimal bottleneck.
The only assumption going into the derivation of~\rf{*2}
is the distribution \rf{*1}
and the approach is well suited for dynamically triangulated surfaces
\cite{david1,adf,david2,kkm}. Here the area of a region
can be identified with the number of triangles in the region
and a  minimal bottleneck is a simple closed path consisting
of three links which partitions the surface into two parts
- the minbu containing $V_2 \gg 1$ triangles, and the parent
containing $N_2 - V_2$ triangles.

Until now the measurement of $\g$ by numerical simulations has been
a rather painful procedure since one had to perform a whole sequence
of simulations (or a so-called grand canonical simulation where the number
of triangles changes, which is also an unpleasant task) for different
areas $N_2$ and in the end fit the measured quantities to a formula related to
\rf{*1}. Following the suggestion in \cite{jm} one can now extract
$\g$ from a single simulation using \rf{*2}. This is neater also
because one does not have to subtract out the leading $e^{\mu_0 N_2}$
piece while making the fit; \rf{*2} contains only the universal
power law piece.
This method also works in practise.
Indeed, in a recent numerical simulation it was possible to extract $\g$ for
pure 2d gravity with high precision (less than 1\%) \cite{ajt}.

\section{The model}

A model of 4d quantum gravity which generalizes the discretized 2d
approach was suggested a year ago \cite{am,aj}. The partition function
is given by
\beq{*3}
Z(\k_2,\k_4)=\sum_{T\in \cT} e^{-\k_4 N_4 +\k_2 N_2}
\eeq
where the sum is over triangulations $T$ in a suitable class of
triangulations $\cT$. The quantity
$N_4$ denotes the number of 4-simplexes in the triangulation and $N_2$ the
number of triangles. The coupling constant
$\k_2$ is inversely proportional to the bare gravitational coupling
constant, while $\k_4$ is related to the bare cosmological constant.
The most important restriction
to be imposed on $\cT$ is that of a fixed topology. If we allow an
unrestricted summation over all topologies in \rf{*3} the partition function
is divergent \cite{aj}. In the following we will always restrict
ourselves to considering only manifolds with the topology of $S^4$.

$Z(\k_2,\k_4)$ is the grand canonical partition function.
It is defined in a region
$\k_4 \geq \k_4^c(\k_2)$ in the $(\k_2,\k_4)$ coupling constant plane.
The only way in which we can hope to obtain a continuum limit is by
letting $\k_4$ approach $\k_4^c(\k_2)$
from above. This tentative continuum limit depends only on
one coupling constant $\k_2$.
We can write \rf{*3} as
\beq{*4}
Z(\k_2,\k_4) = \sum_{N_4} Z(\k_2,N_4) e^{-\k_4 N_4}.
\eeq
$Z(\k_2,N_4)$ is  the canonical partition function
where $N_4$ is kept fixed. Then we have in practise only one coupling
constant, $\k_2$,
and the aspects of gravity which do not involve the fluctuation of the
total volume of the universe can be addressed in the limit of large $N_4$.
However it is actually possible to extract the
critical exponent for volume fluctuations from $Z(\k_2,N_4)$.
Let us assume that the canonical partition function has the
form:
\beq{*5}
Z(\k_2,N_4) = N_4^{\g (\k_2)-3} \; e^{\k_4^c(\k_2) N_4}  \cdot
\left( 1+ \cO (1/N_4)\right)
\eeq
When $\k_4$ is close to $\k_4^c(\k_2)$ we can approximate \rf{*3} with
\beq{*6}
Z(\k_2,\k_4) \sim {\rm analytic}+
\frac{C(\k_2)}{(\k_4-\k_4^c (\k_2))^{\g(\k_2)-2}}
\eeq
where ``analytic'' means possible analytic terms of the form
$(\k_4-\k_4^c)^n$, $0 \leq n < 2-\g(\k_2)$.
 The entropy exponent of the number of 4d simplicial manifolds, $\g$,
determines the volume fluctuations since we have
\beq{*7}
\la N_4^2\ra -\la N_4\ra^2 = \frac{d^2 \ln Z(\k_2,\k_4)}{d\k^2_4}\
\sim \;{\rm analytic} +
\frac{C(\k_2)}{(\k_4-\k_4^c (\k_2))^{\g(\k_2)}}
\eeq
if we assume $\g(\k_2) < 2$.

We will try to extract the critical exponent $\g(\k_2)$ by numerical
simulations. At this point it is important to note that it is by no
means obvious that the exponent $\g$ exists. Equation~\rf{*5} is an
{\it ansatz}, inspired from 2d quantum gravity where  we know that
a similar formula holds when the central charge of matter $c <1$.
In the case of $c =1$ there are logarithmic corrections, and the
form is not known for $c >1$. In 3d simplicial gravity
it was shown \cite{bk,av,av1} that a more likely form
of $Z(\k_1,N_3)$ (the 3d analogue of \rf{*5}) is
\beq{*8}
Z(\k_1,N_3) \sim \exp
\left[ \k_3^c(\k_1) N_3
\left( 1- \frac{C(\k_1)}{N_3^{\a(\k_1)}} \right) \right]
\eeq
and a possible power law correction would be subleading. However,
3d simplicial gravity seems to differ from 4d simplicial gravity
in many respects (\cite{am,aj,abkv}), and it was clear from
the fine tuning process $\k_4 \to \k_4^c(\k_2)$ in the Monte Carlo
simulations that the corrections to the leading term $\exp (\k_4^c N_4)$
were less severe in 4d than in 3d. It is therefore indeed possible that
the finite size corrections in 4d are power-like, as in 2d simplicial
gravity, rather than of the exponential form \rf{*8} encountered in 3d.

\vspace{12pt}

\noindent
Let us now define the minimal bottleneck baby universes, which we,
following \cite{jm}, will denote ``minbu's''. On a 2d triangulated
manifold one can check whether we have a minbu in the following simple
way: pick a link and two neighbouring links, attached to the two vertices
of the first link. If the two links have a vertex in common the three
links form a closed loop. In general this loop will be trivial in the
sense that it will just be the boundary of one of the two triangles in the
surface which contains our starting link. However, it might turn out that
there is no triangle in the surface which has the closed loop as
its boundary. If we cut the surface along such a closed loop we will have
separated it into two disconnected open surfaces each one having
the closed loop as its boundary, provided the topology of the surface
is that of a sphere (which we will assume). Both components will
have spherical topology  when closed by adding the  triangle
which has the loop as boundary. The smallest of the spheres will
be a minbu, the largest the ``mother universe''.

This construction can immediately be generalized to 4d triangulated
manifolds. We check for minbu's as follows: pick a 3-simplex (a tetrahedron)
and in this the four  2-simplexes (triangles) which constitute its boundary.
Identify for each triangle all the 3-simplexes in the manifold
which have this triangle as their boundary.
We now have four groups of 3-simplexes
and all 3-simplexes have one vertex which does not belong to the triangle
from which the 3-simplex was constructed. Pick now a 3-simplex from each of
the 4 groups and check whether their 4 free vertices coincide. If that
is the case the four 3-simplexes together  with the original 3-simplex can
be thought of as the (closed) boundary of a 4-simplex. In general the
4-simplex found by this procedure will just be one of the two 4-simplexes
which shared the original 3-simplex. However, if that is not the case
we will, if the topology of our original 4-manifold is that of a sphere
(which we assume as usual), separate it into two non-trivial components
 if we cut along
the closed piecewise linear 3d manifold built out of the five 3-simplexes.
Both components have this closed 3-manifold as their boundary and
both components
will have the topology of $S^4$ if we close them by
adding to each of them a 4-simplex in such a way that their boundary is
identified with the boundary of the 4-simplex, just with opposite
orientation. The smallest of the 4-spheres will be called the minbu,
the largest the ``mother'', in agreement with the 2d notation.

It is almost clear that the above description can be turned into an
efficient numerical algorithm (after, admittedly, some pain with double
counting etc.) provided one is given the coincidence matrix of the
triangulated 4-manifold.

\vspace{12pt}

Let us now argue that the counting arguments in \cite{jm}
extend to the 4d triangulated manifolds as well.
Let us by $Z'(\k_2,N_4)$ denote the canonical partition function for
4-manifolds consisting of $N_4$ 4-simplices
where one 4-simplex is marked. Generically we will get a different
manifold for each mark so for large $N_4$, where accidental
symmetries are expected to play no role, we get $Z'(\k_2,N_4)
\approx N_4 Z(\k_2,N_4)$.
Such a marked
manifold can also be viewed as a manifold with $(N_4-1)$ 4-simplexes
and a minimal boundary of the kind associated with minbu's
(by removing the interior of the marked 4-simplex).
A moments reflection will convince the reader that the average number of
minbu's of volume $V_4$ on  4-manifolds of total volume $N_4$ will be
given by
\beq{*9a}
\cN_{N_4}(\k_2,V_4) \approx \frac{60}{Z(\k_2,N_4)}  Z'(\k_2,V_4)
Z'(\k_2,N_4-V_4)
\eeq
where $60$ is the number of ways one can glue the two boundaries of
the minbu and the mother together with opposite orientation of
the boundary.

If we {\it assume} the canonical partition function is given by
\rf{*5}, we get
\beq{*9}
\cN_{N_4}(\k_2,V_4) \sim\; C(\k_2)\; N_4 \;V_4^{\g(\k_2)-2}
\eeq

\vspace{12pt}

\noindent
The scene is now set for a numerical determination of the number
of minbu's since it  is already by now
\cite{am,aj,varsted,am1,ajk,bruegmann,bm}  standard how to generate
by Monte Carlo simulations
the class of triangulated 4-manifolds corresponding to the partition
function \rf{*3}. It is also well known \cite{aj} how one can effectively
stay in the neighbourhood of a given volume, $N_4$, even if one
is using the grand canonical partition function \rf{*3}, as one is
forced to by ergodicity requirements in 4d, contrary to what is the case
in 2d.

\section{Numerical results}

By Monte Carlo simulations we have generated a number of 4d-manifolds
according to the distribution dictated by the partition function
\rf{*3}. For details about this procedure we refer e.g. to
\cite{aj,ajk}.
We have considered $\k_2=0.0$, 0.8, 1.0, 1.1 and 1.4 and $N_4=4000$,
9000, 16000\footnote{We have also made a few runs for $N_4=32000$ and
$k_2=1.0$.}. For each value of $\k_2$ and for each value of $N_4$
we have generated approximately 1000 configurations.

Let us at this point review some of the results of \cite{aj}: For
$\k_2 \approx 1.0-1.1$ we observed a transition in geometry from
a highly connected phase with a seemingly large Hausdorff dimension
to a phase with an elongated, almost one-dimensional geometry.
This is why our main efforts have been concerned with
this region of coupling constant space.

The thermalization time is short in the highly connected phase, but quite
long for $\k_2 > 1.0$. In the case of $N_4 =16000$ we have for these
values of the coupling constant used 10000 sweeps\footnote{By a sweep
we mean $N_4$ {\it accepted} updatings.} for thermalization and performed
measurements after each successive fifty sweeps (after each successive
hundred sweeps for $\k_2 =1.4$). The result of the measurements of
minbu's is shown in fig.\ 1 for $N_4=16000$. We have plotted
the number of minbu's on a log-log scale since the formula \rf{*9} suggests
the dependence:
\beq{*10}
\log \cN_{N_4} (\k_2,V_4) = {\rm Const}(\k_2)+ (\g(\k_2)-2)\log V_4
\eeq
The straight lines are the result of a least $\chi^2$ fit.
In fig.\ 2 we have shown the results for $k_2=1.1$ and different values
of $N_4$ and finally in fig.\ 3 we have shown $\g(\k_2)$
as extracted from fig.\ 1.

\vspace{12pt}

\noindent
A few comments should be made. We have only taken into account minbu's with
a  volume $V_4 \geq 9$. For small values of $V_4$ the numbers
$\cN_{N_4} (\k_2,V_4)$ fall into two classes according to
whether $V_4= 4n-1$ or $V_4 = 4n+1$. This is a clear finite size effect.
We have only included in our analysis the last
class of numbers as they seem to fit \rf{*10} all the way down to
$V_4 = 9$. For large volumes $V_4$ of the minbu's the results are average
values for volumes in the neighbourhood of the plotted $V_4$ since the
statistics for a specific large volume is not good.
The errorbars resulting from the binning procedure are smaller than
the symbols used in the figures.

In general we see
a clear deviation from \rf{*10} when $\cN (V_4) < 0.1$ and when
$\k_2 < 1.1$. In this case  the number
of minbu's begins to drop faster than given by \rf{*10}. We have not
shown these data, since they have bad statistics, and we do not know
whether the above mentioned behaviour reflects that \rf{*10} is
not really valid for large $V_4$
or rather that the statistics is just not good enough
(there is typically less than one such large minbu per universe)
for the volumes of manifolds
we consider.  The message we get by comparing
different $N_4$'s does not allow us at the moment to answer this question.
For $\k_2$ larger than 1.1 the situation is different: The number of baby
universes is much larger and even if the number of very large baby
universes is small and errorbars large their numbers seem not to
drop below numbers predicted by \rf{*10}.

It is seen that the curve for $\k_2=1.4$ bends at $V_4 \approx 30-40$
to a different slope. We have used the part with the larger values of
$V_4$ to extract $\g(\k_2)$.
Such a bending is not observed for $\k_2 \leq 1.0$, i.e. in the
phase with a highly connected geometry. A close look at the data for
$\k_2 =1.1$ reveals a slight tendency to such a bending, and since
$\k_2=1.1$ is just at the borderline of the transition to the
elongated geometry the bending is probably related to the fact
that the very fractal structure observed in this phase is only
unfolded for large volumes $V_4$.

\section{Discussion}

With the above reservation concerning the interpretation of the
numerical data, we still consider the results as somewhat remarkable.
It seems possible by numerical simulations to study in detail the
fractal structure of quantum gravity. Indeed, our data contains a
lot of information about the branching of the geometry which will be published
in a more extensive report elsewhere \cite{ajk1}. Let us just here
mention that the two phases of 4d simplicial
quantum gravity observed in former works appear quite clearly in
the minbu measurements.
In the phase with highly connected geometry there is one
``big mother'' which contains the major fraction of the volume
(from 95\% to 80\%  depending on $\k_2$). In the other phase of very
elongated geometry this is not so and there seems genuine democracy
in ``mother size''. {\it Nevertheless the $\g(\k_2)$ extracted seems
perfectly smooth when we pass the transition in geometry}, as is
apparent from fig. 3. It is interesting to note that a similar smoothness
has been observed in the numerical simulations of 2d quantum gravity
\cite{ajt,abk} when we pass the $c=1$ barrier. It points to the
fact that subleading corrections to \rf{*1} might play a very
important geometrical role, and it might be possible to understand
analytically the interplay between the subleading terms in \rf{*1}
and the fractal geometry of quantum gravity. The seed to such an
analysis is already to be found in \cite{jm}.

We would like to stress that the results here, especially for
$\k_2 \leq 1.1$, are obtained for quite small baby universes and we see
some deviations for larger volumes, as mentioned above. It would
clearly be desirable to let the babies grow, but as is well
know from biology this requires time, in our case computer time.
Work in this direction is in progress. It is nevertheless somewhat
intriguing that the transition between the two types of geometry
takes place at a value of $\g \approx -0.5-0.0$, i.e. the same values
of  $\g$ which are of main interest in 2d gravity. Note further that
the results for $\g$ seem internally consistent in the sense the value
in the phase with almost linear geometry (Hausdorff dimension not
larger that two) seems close to the  value $\g=1/2$ for the
so-called branched polymers.
In addition, assuming that the extraction
of $\g$ for $\k_2 <1.1$ will survive the test of large babies, one
could wonder if this reflects that the situation is
like in 2d gravity  where we have a $\g(c)$ depending on the
central charge. In our case $\k_2$ would then play a role
somewhat similar to the central charge. The range of $\g$'s is in agreement
with such a picture. An interesting continuum approach somewhat related to
such an interpretation  can be found in a recent
paper by Antoniadis, Mazur and Mottola \cite{amm}, where the authors
develop a quantum theory for the conformal mode of gravity.

\vspace{24pt}

\newpage

\addtolength{\baselineskip}{-0.20\baselineskip}

\newpage

\addtolength{\baselineskip}{0.20\baselineskip}

\noindent {\large \bf Figure Captions}

\vspace{12pt}

\begin{itemize}
\item[Fig.\ 1] The number of minbu's for $N_4=16000$ and $k_2 =0.0(\Box)$,
$0.8(\bigcirc)$, $1.0(\triangle)$, $1.1(+)$ and $1.4(\times)$.
The straight lines represent the best fits to \rf{*9}.

\item[Fig.\ 2] The number of mimbues for $k_2=1.0$ and $N_4=4000(\Box)$,
$9000(\bigcirc)$, $16000(\triangle)$ and $32000(+)$, normalized such
that $\cN_{N_4} (V)$ is multiplied by $16000/N_4$. (log-log scale)

\item[Fig.\ 3] $\g(\k_2)$ as extracted from fig.1.

\end{itemize}
\end{document}